\documentclass[a4paper,fleqn]{cas-dc}

\usepackage[numbers]{natbib}

\usepackage{tabularx}
\usepackage{booktabs}
\usepackage{array}

\usepackage{enumitem}
\setlist{nosep}

\usepackage{graphicx}
\usepackage{subcaption}

\usepackage{algorithm}
\usepackage{algorithmicx}
\usepackage{algpseudocode}

\begin{document}
\let\WriteBookmarks\relax
\def\floatpagepagefraction{1}
\def\textpagefraction{.001}

\newcommand{\highlight}[1]{\textcolor{blue}{#1}}


\shorttitle{KiseKloset for Fashion Retrieval and Recommendation}    
\shortauthors{Thanh-Tung Phan-Nguyen et al.}  

\title [mode = title]{KiseKloset for Fashion Retrieval and Recommendation} 

\author[1,2]{Thanh-Tung Phan-Nguyen}[orcid=0009-0001-1054-5106] \ead{pnttung@selab.hcmus.edu.vn}

\author[1,2]{Khoi-Nguyen Nguyen-Ngoc}[orcid=0000-0001-7849-7203] \ead{nnknguyen@selab.hcmus.edu.vn}

\author[3]{Tam V. Nguyen}[orcid=0000-0003-0236-7992] \ead{tamnguyen@udayton.edu}

\author[1,2]{Minh-Triet Tran}[orcid=0000-0003-3046-3041] \ead{tmtriet@fit.hcmus.edu.vn}

\author[1,2]{Trung-Nghia Le}[orcid=0000-0002-7363-2610] \ead{ltnghia@fit.hcmus.edu.vn}
\cormark[1]

\cortext[1]{Corresponding author}


\affiliation[1]{organization={University of Science}, 
            city={Ho Chi Minh City},
            country={Viet Nam}}

\affiliation[2]{organization={Vietnam National University}, 
            city={Ho Chi Minh City},
            country={Viet Nam}}
            
\affiliation[3]{organization={University of Dayton}, 
            state={Ohio},
            country={United States}}

\begin{abstract}
The global fashion e-commerce industry has become integral to people's daily lives, leveraging technological advancements to offer personalized shopping experiences, primarily through recommendation systems that enhance customer engagement through personalized suggestions. To improve customers' experience in online shopping, we propose a novel comprehensive KiseKloset system for outfit retrieval and recommendation. We explore two approaches for outfit retrieval: similar item retrieval and text feedback-guided item retrieval. Notably, we introduce a novel transformer architecture designed to recommend complementary items from diverse categories. Furthermore, we enhance the overall performance of the search pipeline by integrating approximate algorithms to optimize the search process. Additionally, addressing the crucial needs of online shoppers, we employ a lightweight yet efficient virtual try-on framework capable of real-time operation, memory efficiency, and maintaining realistic outputs compared to its predecessors. This virtual try-on module empowers users to visualize specific garments on themselves, enhancing the customers' experience and reducing costs associated with damaged items for retailers. We deployed our end-to-end system for online users to test and provide feedback,  enabling us to measure their satisfaction levels. The results of our user study revealed that 84\% of participants found our comprehensive system highly useful, significantly improving their online shopping experience.
\end{abstract}


\maketitle


\section{Introduction}

The fashion industry has been an important part of people's lives. The rise of digital globally, coupled with recent global events such as the pandemic, has reshaped consumer behavior dramatically. E-commerce has surged in popularity and sophistication, with even traditional retail outlets expanding their presence on online platforms, such as Amazon and Zalando, etc. Consequently, it is hardly surprising that the fashion e-commerce market will continue to flourish and diversify at an extraordinary pace. The global fashion e-commerce market is expected to reach a value of over 820 billion U.S. dollars in 2023 and could be 1.2 trillion U.S. dollars in 2027~\cite{Fashion2327-Statista2023}. 
Nevertheless, the shift from traditional in-store shopping to online shopping presents distinctive challenges. These include the time-consuming process of item selection without the aid of sales staff recommendations, limitations in trying on clothes before making a purchase, and risks and costs associated with product transportation.

In response to these challenges, intelligent fashion technologies have emerged as promising solutions, aiming to replicate the support and comfort of in-person shopping experiences. Among them, three significant areas of intelligent fashion have drawn significant attention from both academia and industry communities: outfit retrieval and recommendation (ORR), and virtual try-on (VTON) systems. These technologies share a common objective of augmenting the customer experience on online platforms, thereby enabling retailers to bolster profitability. ORR systems~\cite{Lin-CVPR2020-Fashion, Baldrati-CVPR2022-Conditioned} facilitate easier and smarter shopping choices by recommending items that align with users' preferences and style. Meanwhile, VTON allows users to virtually wear garments, enriching the customer experience and saving the cost associated with damaged items for retailers. These solutions redefine the landscape of online shopping and hold the promise of elevating the customer experience to the next level.


ORR has always been crucial to modern fashion e-commerce systems, offering numerous benefits, such as boosting conversion rates by up to 300\% and increasing revenue by up to 31\%~\cite{web-recommendation-motivation}. Instead of spending hours searching for fashion items online or in physical stores, users can rely on these systems to discover new and relevant clothing, saving their valuable time and effort. Moreover, these systems encourage users to explore a wider range of options, exposing them to different styles or brands. This encourages users to try out something they might not have considered before, thus improving the fashion industry's overall revenue. Meanwhile, VTON helps customers visualize how a garment might look on their body based on images of themselves and the desired item. This technology addresses a substantial challenge online shoppers encounter: having to visit a store to try on clothing items physically. 

\begin{figure}[t!]
  \centering
  \includegraphics[width=\linewidth]{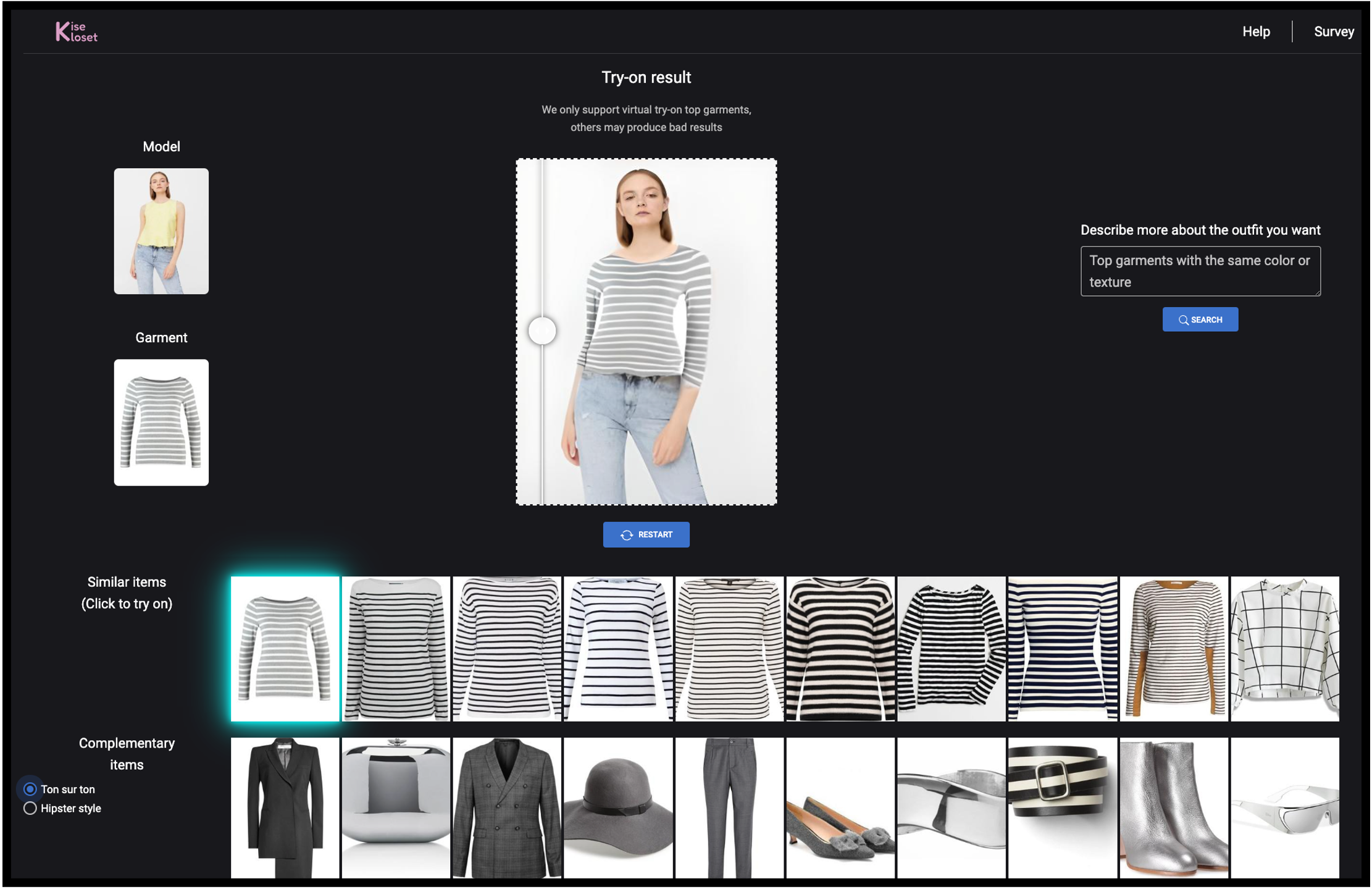}
  \caption{Interface of the proposed KiseKloset system, integrated outfit retrieval, recommendation, and virtual try-on capabilities.}
  \label{fig:KiseKloset}
  \vspace{-5mm}
\end{figure}

This paper aims to address significant challenges in fashion e-commerce, including ORR and VTON. Leveraging cutting-edge technologies, we develop a novel comprehensive system, namely KiseKloset, that enhances the online shopping experience and bridges the gap between the physical and digital realms of fashion retail. Particularly, our KiseKloset system supports users in finding their desired outfits via intra-category similar item retrieval and text feedback-guided item retrieval (\autoref{fig:KiseKloset}). Additionally, we propose a novel transformer network for inter-category complementary item recommendation. Taking a fashion item as the reference item, we recommend items in a complete outfit that can match with the reference item based on its visual features. Unlike previous works~\cite{Mariya-ECCV18-Learning, Sarkar-CVPRW2022-OutfitTransformer}, which could only recommend 
 a single missing item, our proposed method is more generalized, allowing for varying numbers of missing items in an outfit. We also incorporate approximate searching methods to enhance the system's overall performance. To address the issue of the complexity of existing VTON methods, we employ a lightweight teacher-student network that offers faster and more resource-efficient performance while ensuring the fidelity of the outcomes \cite{Nguyen-ISMAR2023-DMVTON}. 

 We conducted experiments to evaluate approximate search algorithms on our proposed framework on large-scale fashion datasets. The results show that our system can achieve scalable performance in terms of speed, accuracy, and memory usage. These findings demonstrate that KiseKloset holds significant practical potential, offering new possibilities for personalized and immersive interactions between consumers and online fashion platforms. In addition, we deployed the KiseKloset system for trial usage to evaluate its effectiveness and measure user satisfaction. Online users were invited to test the end-to-end system and provide feedback on their experience. The results of the user study indicate an overall positive reception, with 84.4\% of participants reporting that the system was highly useful in enhancing and streamlining their online shopping experience. Beyond direct usability, the study also highlighted strong user advocacy: 65.6\% of users expressed their willingness to recommend KiseKloset to friends and family, underscoring the system’s potential for widespread adoption. These findings demonstrate that KiseKloset not only delivers functional value but also fosters user trust and satisfaction, reinforcing its promise as a practical solution for improving digital shopping journeys. 

The main contribution of this paper can be summarized as follows:
\begin{itemize}
    \item We develop a comprehensive system called KiseKloset, integrating both ORR capabilities and VTON functionality, thereby enhancing the overall user experience and increasing customer satisfaction in the fashion e-commerce sector. As demonstrated in \autoref{tab:functionality_comparison}, our proposed KiseKloset stands as the first system supporting both ORR and VTON for real-word applications.
    \item We investigate different methodologies for searching fashion items from imagery and textual queries. Our approximating search algorithm operates significantly faster than exhaustive methods.
    \item We propose a novel generalized transformer-based outfit recommendation network to address the recommendation of multiple complementary items in an outfit.
    \item We deploy our KiseKloset system for trial usage to measure user satisfaction. The user study indicates that users have an overall positive experience with our system. It also reveals that 84\% of users found our system highly useful and effectively enhancing their online shopping experience.
\end{itemize}

\begin{table*}[t!]
    \centering
    \caption{Functionality comparison of existing ORR and VTON systems. \textit{Outfit} stands for the entire item set. `-' denotes unavailable feature.}
    \label{tab:functionality_comparison}
    
    \resizebox{1\textwidth}{!}{
    \begin{tabular}{l|c|c|c|c}
    \toprule
        \textbf{Method} & \textbf{Year} & \textbf{Outfit Retrieval} & \textbf{Outfit Recommendation} & \textbf{Outfit Try-On} \\ \midrule
         Zhou et al.~\cite{zhou2012image} & 2012 & - & - &  Virtual fitting \\
         Wu et al.~\cite{wu2022design} & 2022 & - & - &  Virtual fitting \\ 
         MagicCloset~\cite{liu2012hi} & 2012 & - & Occasion-oriented item pairing & Copy-Paste \\
         Tangseng et al.~\cite{tangseng2017recommending} & 2017 & - & Outfit generation & - \\
         Alibaba iFashion~\cite{chen2019pog} & 2019 & - & History-based outfit generation & - \\
         Jang et al.~\cite{jang2024lost} & 2024 & - & Textual event-based outfit generation & - \\
        \rowcolor{lightgray} KiseKloset & This work & Similar and Text feedback-guided retrieval &  Multiple complementary items recommendation & Real-time VTON \\
        \bottomrule
    \end{tabular}
    }
    
    \vspace{-5mm}
\end{table*}

\section{Related Work}

\subsection{Outfit Retrieval And Recommendation}

OOR systems have proven useful by assisting users in discovering relevant and new styles, saving time, and making informed fashion choices. The explosion of large-scale datasets and deep learning has resulted in a number of studies focusing on visual-based ORR. 

\begin{figure}[t!]
    \centering
    \begin{subfigure}{\linewidth}
    \includegraphics[width=\linewidth]{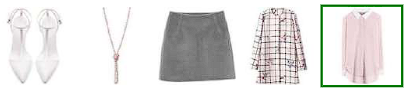}
    \caption{Fill-in-the-blank outfit recommendation~\cite{Mariya-ECCV18-Learning, Sarkar-CVPRW2022-OutfitTransformer}.}
    \label{fig:fitb-task-a}
    \end{subfigure}
    \begin{subfigure}{\linewidth}
    \includegraphics[width=\linewidth]{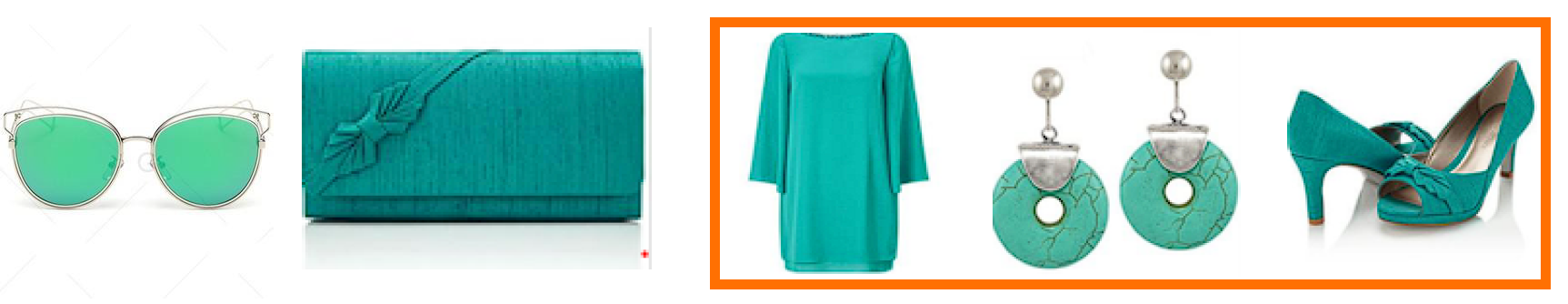}
    \caption{Our inter-category complementary item recommendation.}
    \label{fig:fitb-task-b}
    \end{subfigure}
    \caption{Our complementary item recommendation is more generalized than fill-in-the-blank outfit recommendation~\cite{Mariya-ECCV18-Learning, Sarkar-CVPRW2022-OutfitTransformer}, allowing various number of missing items. Highlighted item is the missing item to be recommended.}
    \label{fig:fitb-task}
    \vspace{-5mm}
\end{figure}

FashionIQ dataset~\cite{Wu-CVPR2021-FashionIQ} was introduced for the text feedback-guided item retrieval task, where we need to find the desired item given a reference input item and its relative attribute feedback in the natural language. PolyvoreOutfits dataset~\cite{Mariya-ECCV18-Learning} was provided to support the fill-in-the-blank task, where we need to find an item of a specific category to complete a partial outfit (see \autoref{fig:fitb-task}). 

Mariya et al.~\cite{Mariya-ECCV18-Learning} learned a shared embedding space where all fashion items are represented and then projected the embedding space into subspaces based on pairs of item types to find items. Lin et al.~\cite{Lin-CVPR2020-Fashion} represented each category by an embedding vector, which is used to predict the attention weights of subspace embeddings created from image embedding via a self-attention mechanism. Attention weights determine the contribution of each subspace embedding in the final output embedding. To eliminate the learning of pair-wise embeddings, Sarkar et al.~\cite{Sarkar-CVPRW2022-OutfitTransformer} learned the embedding of an entire outfit using a transformer encoder and a special BERT-like token~\cite{Devlin-ArXiv2018-BERT}. This token not only captures the embeddings of all items in the outfit but also specifies the target category for the missing item. On the other hand, Baldrati et al. \cite{Baldrati-CVPR2022-Effective, Baldrati-CVPR2022-Conditioned} utilized CLIP~\cite{Radford-OpenAIblog2019-Language} to achieve state-of-the-art performance on text feedback-guided image retrieval. Han et al.~\cite{Han-ECCV2022-FashionViL} bridged the gap between the text feedback-guided image retrieval problem and the fill-in-the-blank problem by proposing a general vision-language network for multiple-task learning, resulting in universal embeddings for downstream tasks. 

Our proposed system supports both text feedback-guided item retrieval and fill-in-the-blank recommendation. However, our method is more generalized than previous works~\cite{Mariya-ECCV18-Learning, Sarkar-CVPRW2022-OutfitTransformer}, enabling the recommendation of different numbers of missing items in an outfit.
 
\subsection{Similarity Search}

Similarity search has become a hot topic in recent years due to the explosion of data. There are two main types of similarity search: exhaustive and approximate. Exhaustive searching, such as the K-Nearest Neighbor (KNN) algorithm, involves calculating the distances between the query point and all items in a dataset to find the nearest neighbours. This approach provides accurate results but can be computationally expensive, especially in large datasets or high dimensional data points.

Approximate searching rises as the solution for the time complexity of exhaustive searching. These methods either reduce the dataset dimension or limit the search scope with the trade-off for accuracy. Production Quantization (PQ)~\cite{Jegou-TPAMI2010-Product} clusters each dimension or group of dimensions into buckets and indexes the dataset by those buckets instead, thus reducing the number of bytes required to represent the dataset. Inverted File Index (IVF)~\cite{Sivic-ICCV2003-Video} divides the search space into subspaces and only searches in a subset of these subspaces. Some methods leverage the tree or graph data structure to speed up the search process with the cost of more memory consumption~\cite{Malkov-TPAMI2018-Efficient}. Specifically, Hierarchical Navigable Small Worlds (HNSW)~\cite{Malkov-TPAMI2018-Efficient} constructs multiple subgraphs from all items; the lower the level is, the denser the subgraph becomes, and the searching process is performed like in a skip list. Approximate Nearest Neighbors Oh Yeah (ANNOY)\footnote{\url{https://pypi.org/project/annoy/}}, used in Spotify's recommendation system, builds a binary tree by repeatedly choosing two random items and splitting the space into two subspaces.

\section{Proposed System}

\subsection{Outfit Retrieval}

\begin{figure}[t!]
    \centering
    \begin{subfigure}{\linewidth}
    \includegraphics[width=\linewidth]{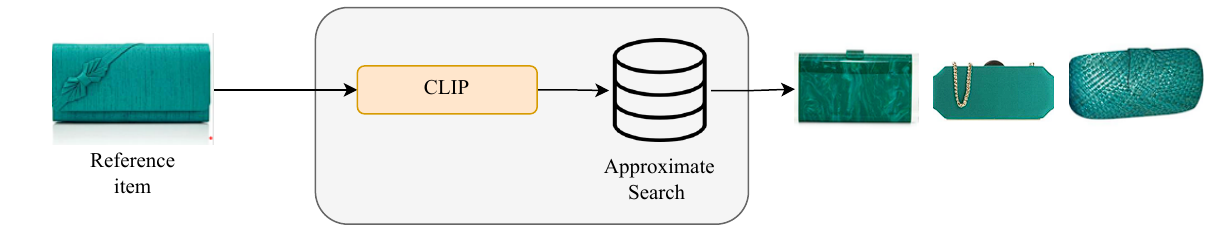}
    \caption{Intra-category similar item retrieval.}
    \label{fig:chapter4-overview-a}
    \end{subfigure}

    \begin{subfigure}{\linewidth}
    \includegraphics[width=\linewidth]{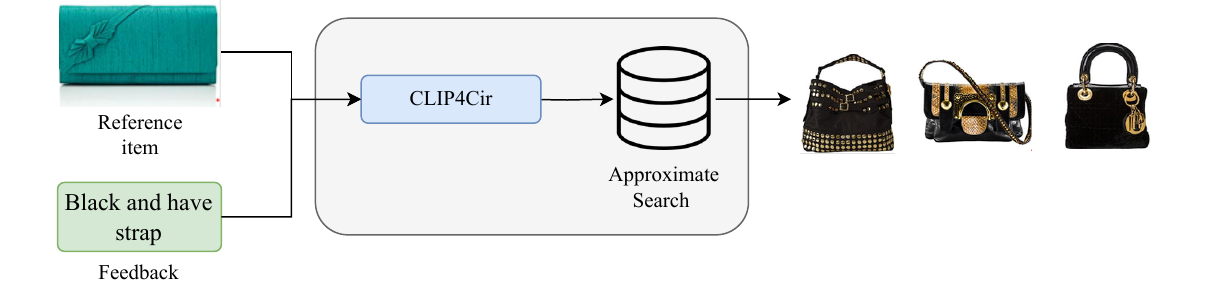}
    \caption{Text feedback-guided item retrieval.}
    \label{fig:chapter4-overview-b}
    \end{subfigure}

    \begin{subfigure}{\linewidth}
    \includegraphics[width=\linewidth]{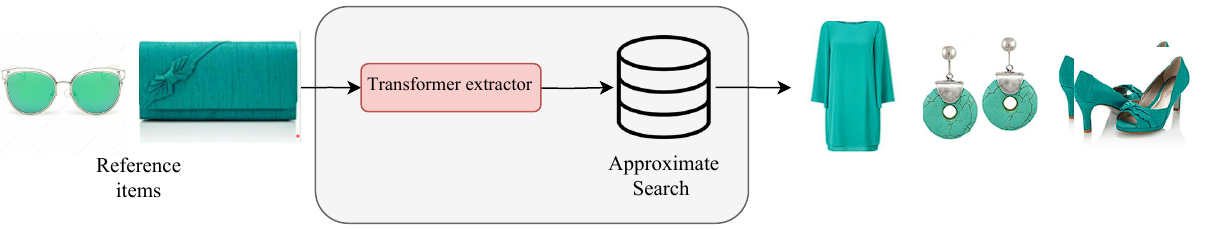}
    \caption{Inter-category complementary item recommendation.}
    \label{fig:chapter4-overview-c}
    \end{subfigure}
    
    \caption{Our system supports various types of ORR.}
    \label{fig:chapter4-overview}
    \vspace{-5mm}
\end{figure}

\autoref{fig:chapter4-overview-a} provides an overview of our \textit{intra-category similar item retrieval} pipeline. Initially, we extract the visual embeddings of the reference image using the same CLIP model~\cite{Radford-OpenAIblog2019-Language} for extracting embeddings across the database. Subsequently, an approximate searching algorithm utilizes these extracted features to retrieve similar items.

\begin{figure}[t!]
    \centering
    \includegraphics[width=\linewidth]{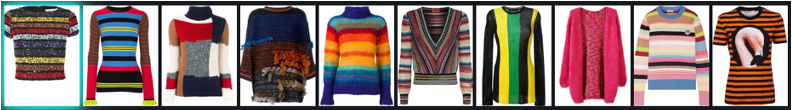}
    User feedback: \textit{Colorful top garments with similar texture}.
    \includegraphics[width=\linewidth]{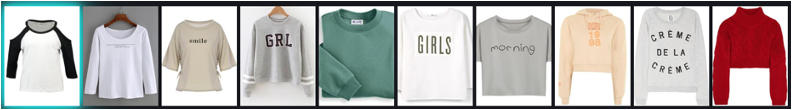}
    User feedback: \textit{Top garments with same color but having text}.
    \caption{Examples of text feedback-guided item retrieval. The first item is reference, the remain items are retrieval results.}
    \label{fig:feedback_retrieval}
    \vspace{-3mm}
\end{figure}

As depicted in \autoref{fig:chapter4-overview-b}, to tackle \textit{text feedback-guided item retrieval}, we adopt the simple yet efficient CLIP4Cir architecture~\cite{Baldrati-CVPR2022-Conditioned}. It utilizes the CLIP model~\cite{Radford-OpenAIblog2019-Language} to extract embeddings from both the reference image and the accompanying textual feedback. These embeddings are then merged into a unified output embedding via a combiner network~\cite{Baldrati-CVPR2022-Conditioned}. The combiner is trained to maximize the similarity between its output embeddings and the embeddings of target images produced by CLIP, using a symmetric contrastive loss. Additionally, we leverage the approximate searching algorithm to retrieve similar items for the resultant output embeddings. Examples of our text feedback-guided item retrieval are illustrated in \autoref{fig:feedback_retrieval}.

\begin{figure}[t!]
    \centering
    \includegraphics[width=\linewidth]{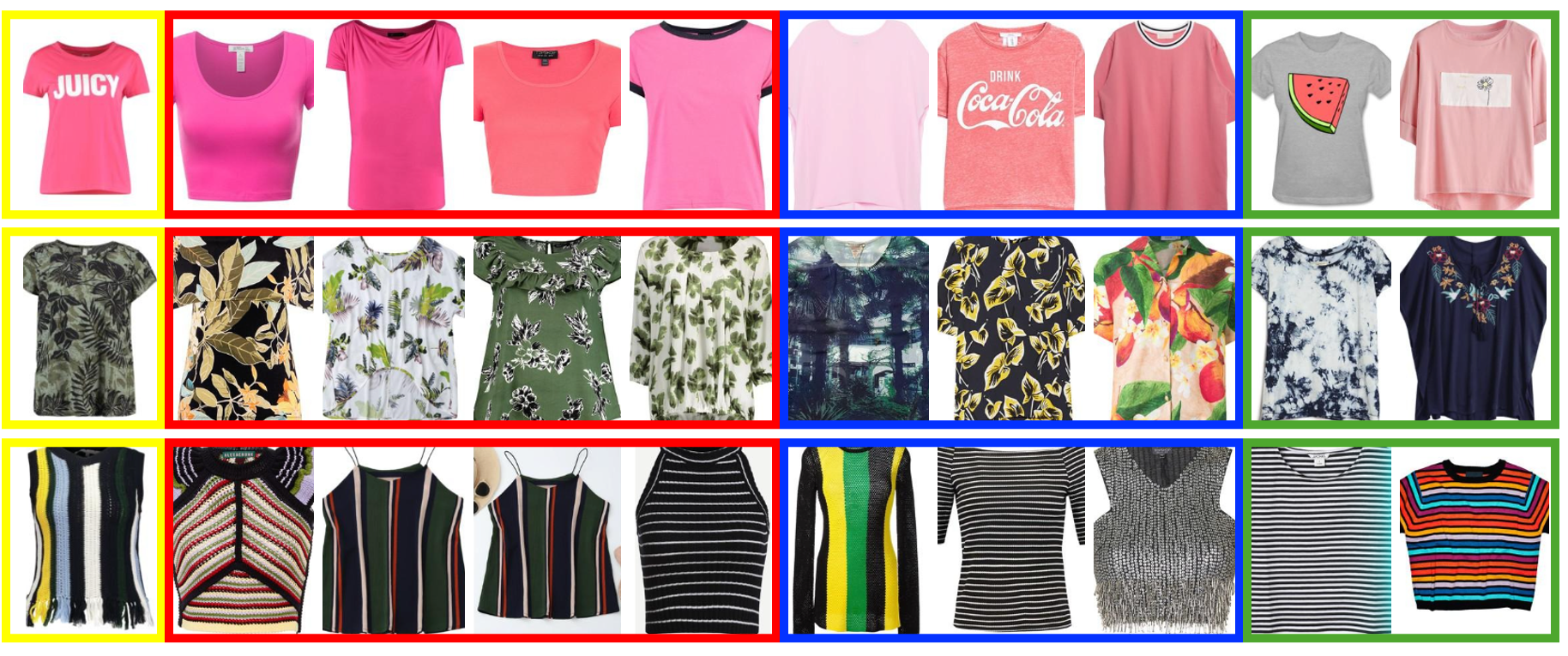}
    \caption{We augment retrieval results to enhance users' experience. \textcolor{yellow}{Yellow}, \textcolor{red}{red}, \textcolor{blue}{blue}, and \textcolor{olive}{green} boundaries stand for reference items, accurately retrieved items, approximately retrieved items, heuristically retrieved items, respectively.}
    \label{fig:augmented_retrieval_results}
    \vspace{-5mm}
\end{figure}

In contrast to traditional retrieval tasks that often yield identical results, the uniformity of fashion retrieval results may diminish user engagement. Thus, to enrich users' experiences, we augment the items in the retrieval results. Specifically, we incrementally introduce variation in the retrieved items (\autoref{fig:augmented_retrieval_results}) as follows:
\begin{itemize}
    \item Accurate retrieval: Retaining the original retrieval results for the first half of the items.
    \item Approximate retrieval: Randomly shuffling items between the top-10 and top-100 retrieval results for the subsequent quarter of items.
    \item Heuristic retrieval: Randomly shuffling items between the top-500 and top-1000 retrieval results for the remaining quarter of items.
\end{itemize}

\subsection{Outfit Recommendation}

We address the inter-category complementary item recommendation task as the sentence generation task in the natural language domain. We investigate using the transformer architecture to retrieve complementary items and make a complete outfit (see \autoref{fig:chapter4-overview-c}). Specifically, we consider a complete outfit as a sequence of fashion items, where the order is determined by their categories: bags, tops, outerwear, hats, bottoms, scarves, jewelry, accessories, shoes, and sunglasses. Not all categories are required. All items in an outfit are divided into three groups: \textit{Input}, \textit{Output}, and \textit{Unavailable}. The division of items among these groups depends on whether it's the training or inference phase.

In the training phase, we divide items into an outfit as follows:
\begin{itemize}
    \item Input items are selected randomly from the available items.
    \item The remaining items within available items are outfit items.
    \item Items in categories that the outfit lacks are denoted as unavailable.
\end{itemize}

During the inference phase, as we generate complementary items for the reference item, we use the following division:
\begin{itemize}
    \item Input item is the reference item only.
    \item Output items belong to categories that we aim to propose to users. These items are our target complementary items.
    \item Items in the remaining categories are denoted as unavailable.
\end{itemize}

Each group of items has different ways to represent their embedding:
\begin{itemize}
    \item Input: the item's visual embedding as-is.
    \item Output: a shared, learnable embedding, denoted as <OUT>.
    \item Unavailable: a shared, learnable embedding, denoted as <UN>.
\end{itemize}

\begin{figure}[t!]
    \centering
    \begin{subfigure}{0.85\linewidth}
    \includegraphics[width=\linewidth]{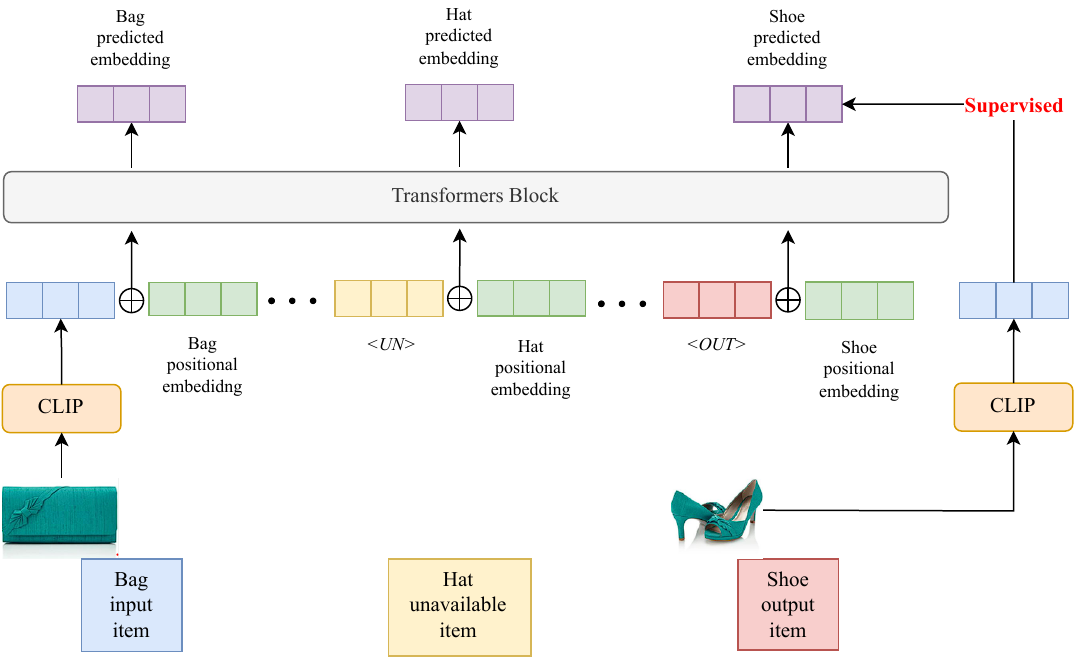}
    \caption{Training phase.}
    \label{fig:chapter4-outfit-trans-a}
    \end{subfigure}

    \begin{subfigure}{0.85\linewidth}
    \includegraphics[width=\linewidth]{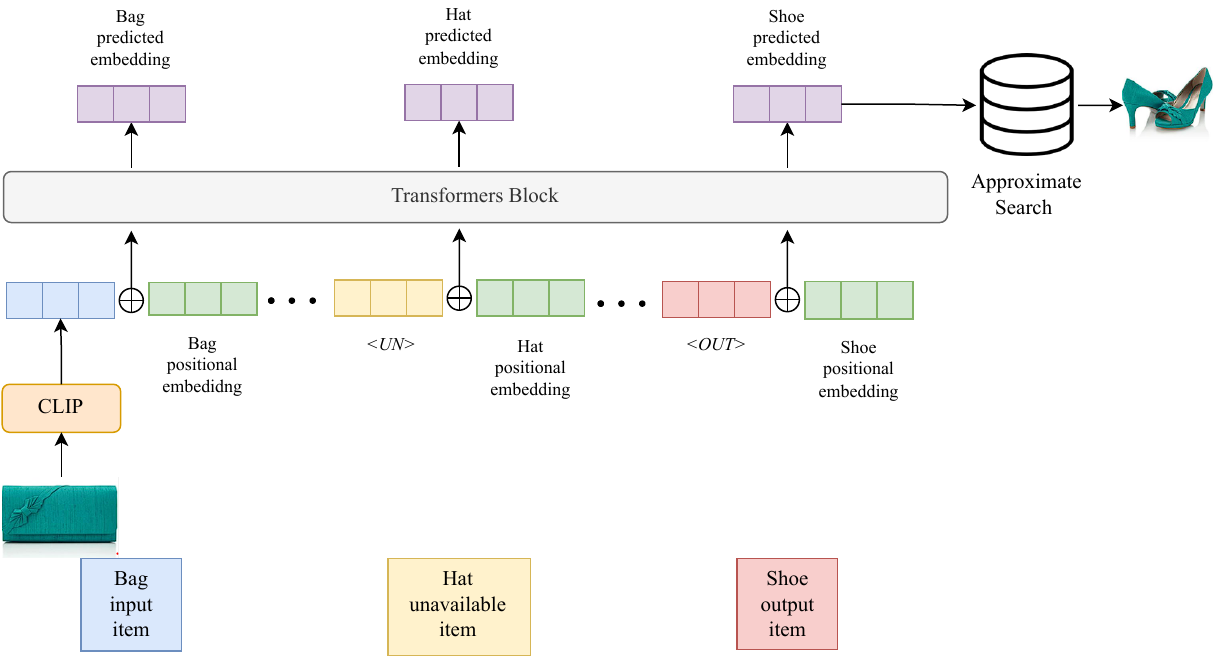}
    \caption{Inference phase.}
    \label{fig:chapter4-outfit-trans-b}
    \end{subfigure}
    
    \caption{Our proposed inter-category complementary item recommendation pipeline.}
    \label{fig:chapter4-outfit-trans}
    \vspace{-3mm}
\end{figure}

All embeddings are fed into a transformers encoder block~\cite{Vaswani-NeurIPS2017-Attention} to produce cross-semantic-aware output embeddings (\autoref{fig:chapter4-outfit-trans}). We also use the positional embeddings technique~\cite{Vaswani-NeurIPS2017-Attention}, where each position specifies the item's category. During the training phase, the network minimizes the noise contrastive loss~\cite{Lai-Arxiv2019-Contrastive} so the transformers block learns to maximize the similarity between its output embeddings and the \textit{output} items' original visual embeddings (as shown in \autoref{fig:chapter4-outfit-trans-a}). We formulate the loss $L$ as follows:
\begin{align}
    & L = -\frac{1}{N}*\Sigma_{i=1}^{N} log{\frac{exp(S_{i_P})}{exp(S_{i_P}) + exp(S_{i_N})}},\\
    & S_{i_P} = cos(pred_i, pos_i),\\
    & S_{i_N} = \Sigma_{j \in {neg_i}} cos(pred_i, j),
\end{align}
where $N$ is the number of \textit{output} items in all outfits, $pred_i$ is the $i^{th}$ predicted output embeddings, $pos_i$ is the ground-truth visual embeddings of that \textit{output} item, $neg_i$ is a set of negative samples which contains the embeddings of other items of the same category.

During the inference phase, we utilize the approximate searching algorithm to retrieve similar items from output embeddings (as shown in \autoref{fig:chapter4-outfit-trans-b}).

\begin{figure}[t!]
    \centering
\includegraphics[width=\linewidth]{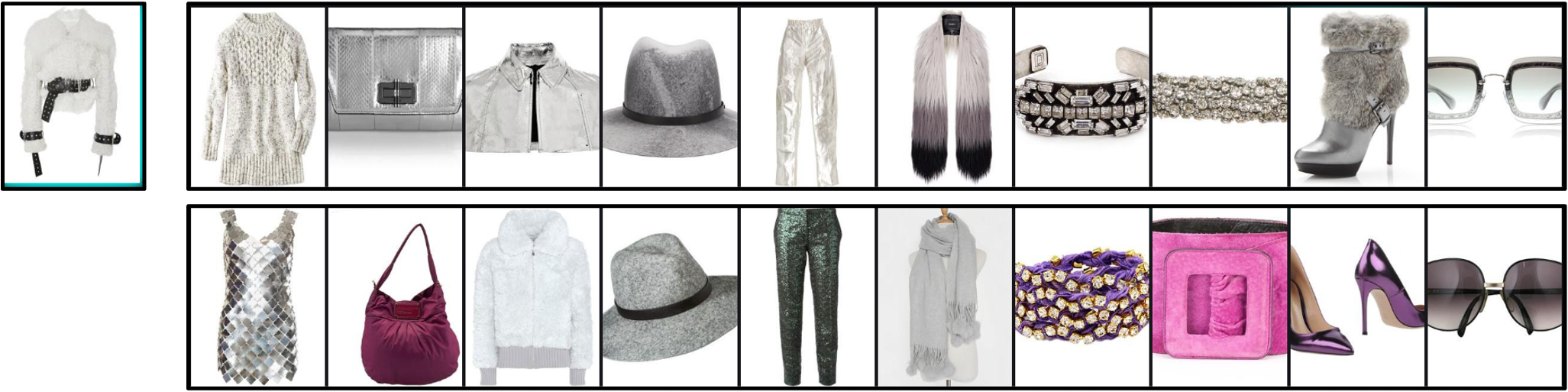} 
    \hfill
    \vspace{0.1mm}
\includegraphics[width=\linewidth]{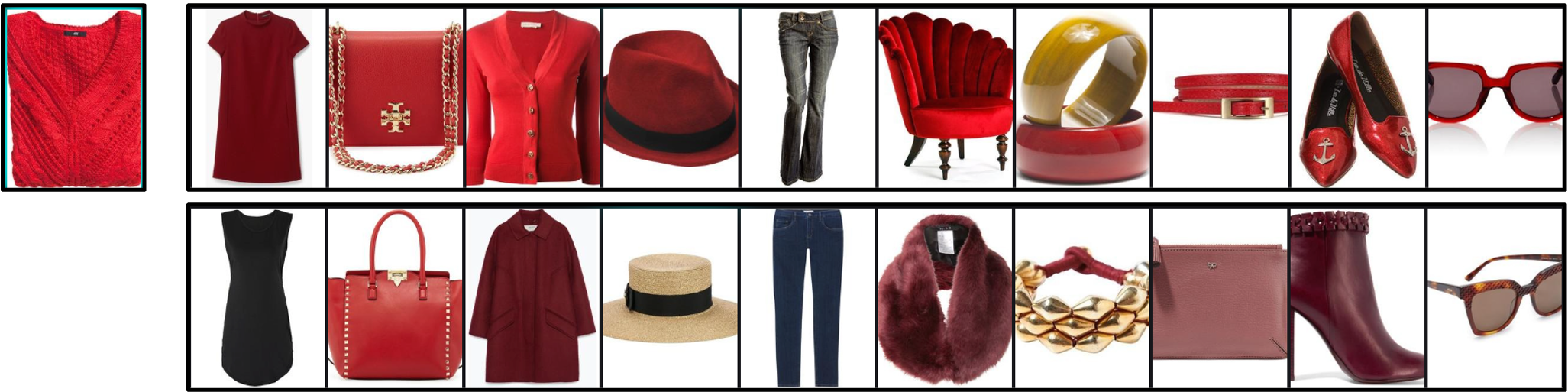} 
    \caption{Two settings of inter-category complementary item recommendation: Tone sur tone (top) and Mix and match (bottom).}
    \label{fig:complementary_recommendation_settings}
    \vspace{-5mm}
\end{figure}

Furthermore, we introduce two settings for complementary item recommendation (\autoref{fig:complementary_recommendation_settings}):
\begin{itemize}
    \item Tone sur tone: Recommended items exhibit similar colors or textures to the query item, easily attainable via top-1 results.
    \item Mix and match (i.e., Hipster fashion style): To inject more vibrancy and diversity into the outfit, we blend output items that may vary in style, color, or texture. This is achieved by randomly shuffling the top-1000 results. 
\end{itemize}

\textbf{Implementation details. }Our proposed transformer block has 6 layers, 8 attention heads, and 11 inputs in total, and each input has 640 dimensions (same as CLIP output dimension). We trained the network from scratch on the PolyvoreOutfits dataset~\cite{Mariya-ECCV18-Learning}. The AdamW optimizer was used with a learning rate of 0.0001 and a weight decay of 0.3. The learning rate scheduler had a step size of 20 with a gamma factor of 0.5.

\subsection{Outfit Try-On}


\begin{figure}[t!]
\centering
\begin{subfigure}{0.45\linewidth}
    \includegraphics[width=\linewidth]{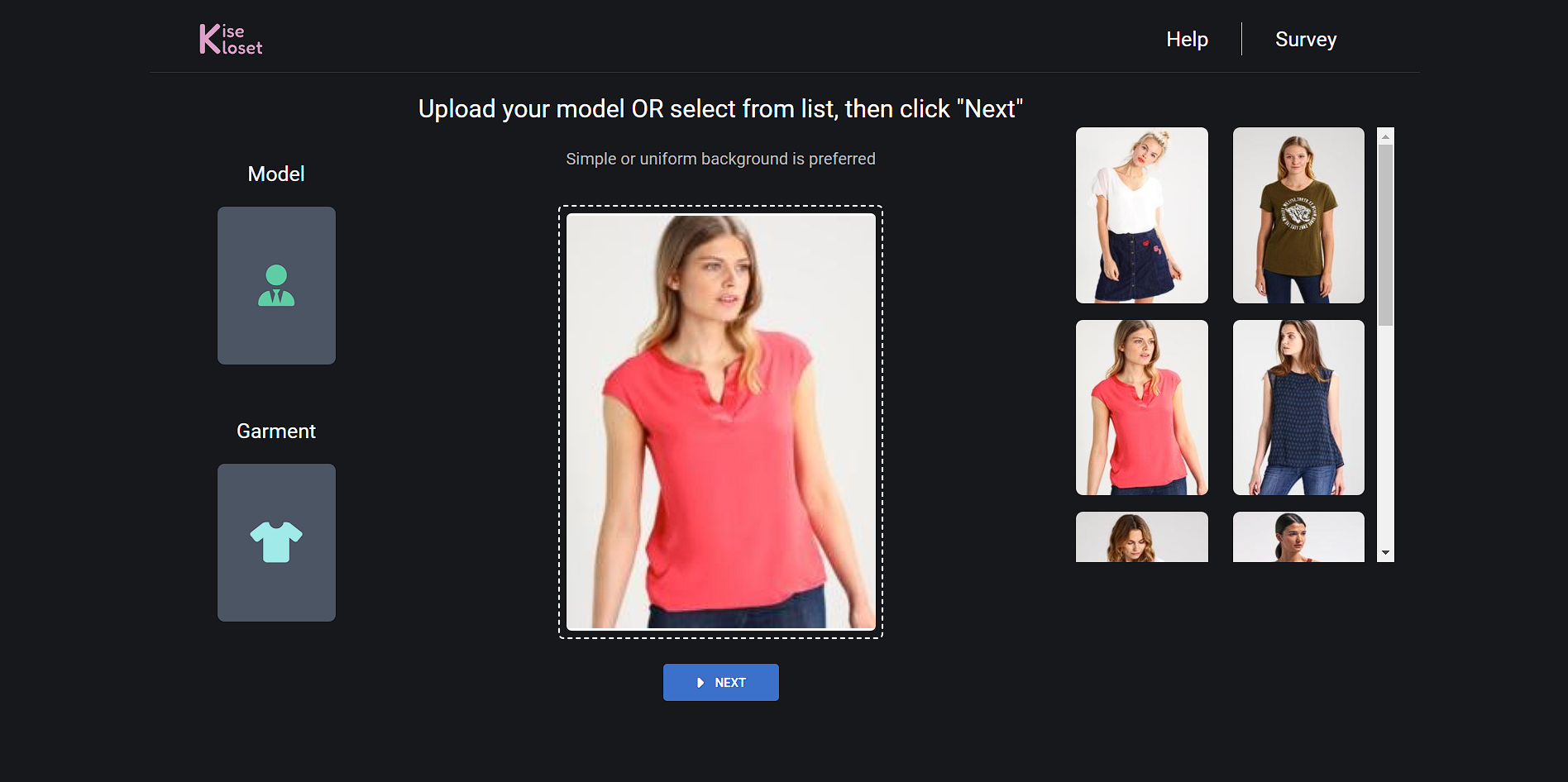}
    \caption{Choose the input person to try on. An example where a user picks one of our provided models}
    \label{fig:web-step1}
\end{subfigure}
\hfill
\begin{subfigure}{0.45\linewidth}
    \includegraphics[width=\linewidth]{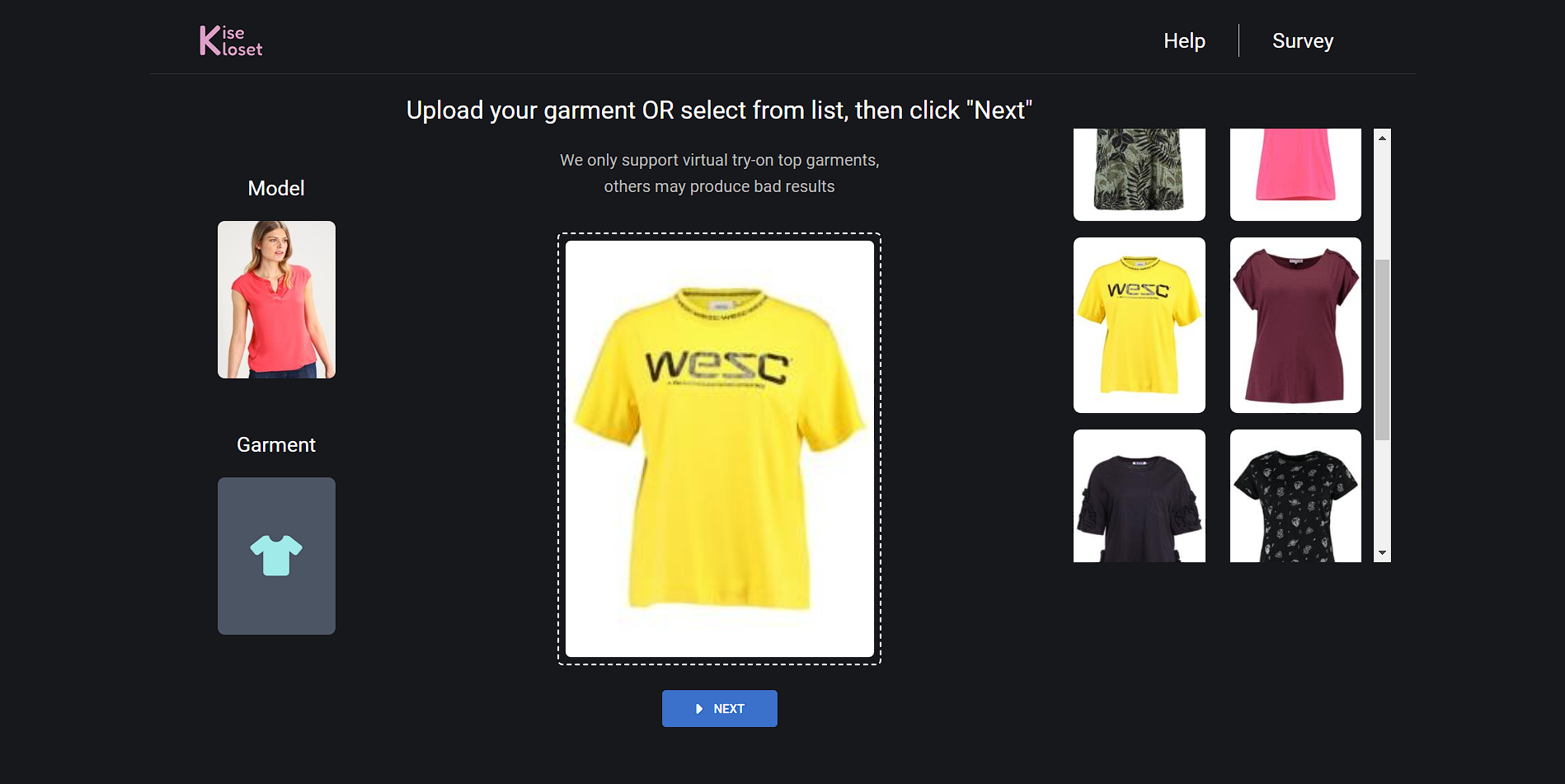}
    \caption{Choose the target garment to try on. Users must pick one of the garments provided on the right panel}
    \label{fig:web-step2}
\end{subfigure}
\hfill
\begin{subfigure}{0.45\linewidth}
    \includegraphics[width=\linewidth]{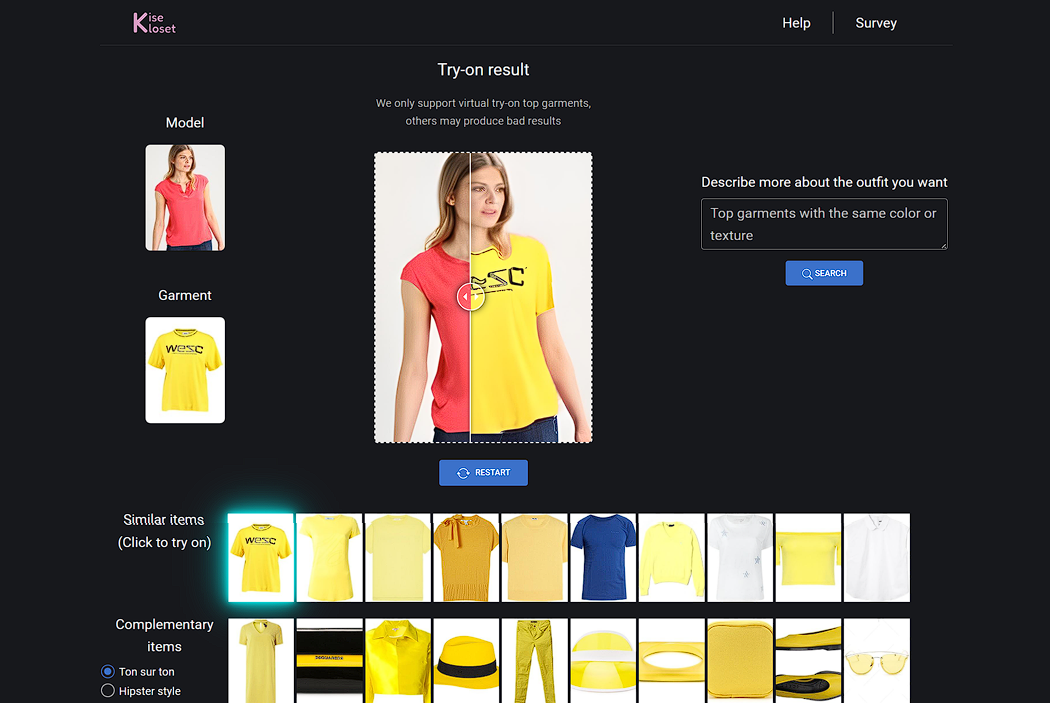}
    \caption{Try-on result and recommendation playground.}
    \label{fig:web-recommend}
\end{subfigure}
\caption{Interaction flow of the proposed KiseKloset system.}
\label{fig:web-flow}
\vspace{-5mm}
\end{figure}

Our objective is to generate an image of a person wearing a specific garment in a real-time manner for practical usage. To achieve this goal, we adopt DM-VTON~\cite{Nguyen-ISMAR2023-DMVTON}, which consists of two networks: teacher and student networks. The teacher network aims to produce the VTON result using the parser-based training process. The student network then utilizes the teacher network to generate synthetic input images, enabling the student network to be supervised by the original images without relying on human representation. The teacher also helps this process through a knowledge distillation scheme. The teacher network is built upon FS-VTON \cite{He-CVPR2022-Style} to ensure high-quality output. Focusing on inference speed, the student network is based on MobileNetV2~\cite{Sandler-CVPR2018-Mobilenetv2} with inverted residual blocks specifically designed to optimize computational efficiency and model size.

\subsection{Interface of KiseKloset System}

 We demonstrate the scenario where users find a particular garment that catches their eye while shopping online at home. They consider buying it but are unsure whether it looks good on them. We facilitate it by offering the KiseKloset system. Using this system, users can try on such garments before purchasing. Upon try-on, the system recommends other fashion items that may suit their tastes and preferences. Users can also find new garments with some relative attributes (i.e. green, longer sleeves) to the selected garment by providing feedback in natural language. \autoref{fig:web-flow} illustrates how our system works.

First, we prepare some human models beforehand as references for users. Users can use those models or upload their photos as input images for the try-on framework. \autoref{fig:web-step1} demonstrates the case where a user picks one of our provided models. After that, users can press the Next button below to advance to the next step.

Subsequently, users must upload or pick the garment they want to try on from our provided list on the right side (as shown in~\autoref{fig:web-step2}). The left panel also displays the chosen model image in the previous step. Now users can press the Next button to view the try-on result. 

Once users have chosen both the input person and garment image, they are presented with the try-on result, as shown in \autoref{fig:web-recommend}. A slider in the middle allows users to compare the original and try-on images. Beneath the try-on result, we provide recommendations for the chosen garment. These recommendations include items within the same category as the reference item and items from other categories that may catch users' attention. Users can try on these recommended items by clicking on them. As for items from other categories, we provided 2 options: Tone sur tone, items with the same color or textures, and, Mix and match, items with mixed style and color. In case users want garments not included in the recommendation list but have some relation to the reference item, we offer a text box on the right side. Users can input their preferences; consequently, we provide a new search based on their feedback.

\begin{figure}[t!]
    \centering
\includegraphics[width=\linewidth]{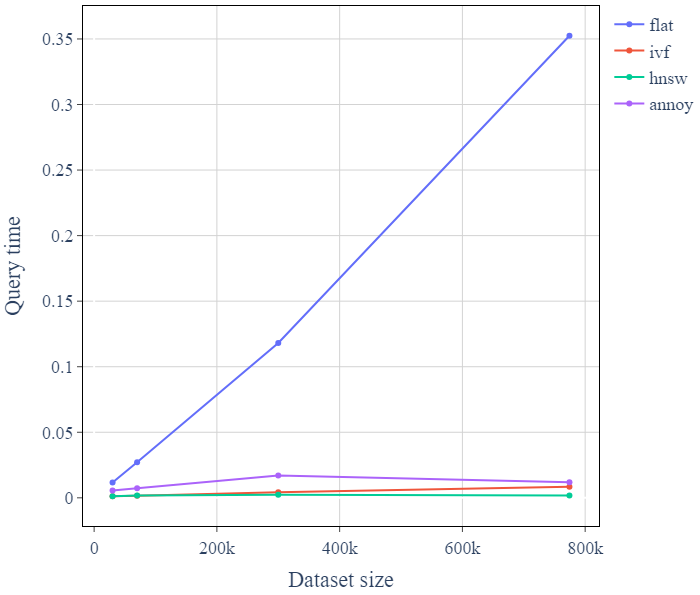}
    \caption{Query time of nearest neighbor methods.}
    \label{fig:chapter4-time-complex}
    \vspace{-5mm}
\end{figure}

\section{Experiments}

\subsection{Ablation Study}

In this ablation study, we aim to demonstrate the effectiveness of used approximate searching algorithms. Particularly, we compare approximate algorithms (e.g., IVF~\cite{johnson-bigdata2019-faiss}, HNSW~\cite{johnson-bigdata2019-faiss}, ANNOY\footnote{\url{https://pypi.org/project/annoy/}}) with exhaustive KNN algorithm.


First, we evaluated the time complexity between KNN, IVF, ANNOY, and HNSW algorithms. We subsequently selected 30.000, 70.000, 300.000, and all items from the merged dataset (PolyvoreOutfits~\cite{Mariya-ECCV18-Learning} and DeepFashion~\cite{Liu-CVPR2016-DeepFashion}) to run those algorithms. \autoref{fig:chapter4-time-complex} illustrates the result, showing that all investigated approximate methods run significantly faster than KNN and have nearly constant runtime. This experiment has proven the scalability of approximate searching methods.

Additionally, we evaluated the query time, recall score and memory usage of methods. \autoref{table:chapter4-ann-all} indicates that HNSW outperforms all other methods regarding query time while using nearly the same memory as KNN and maintaining a recall score of 0.98. These experimental results show the potential of the approximate search in the real-world applications.

\begin{table}[t!]
\centering
\caption{Performance of searching algorithms.}
\begin{tabular}{lrrr}
\toprule
\textbf{Method} & \textbf{Query time (ms)} & \textbf{Recall (\%)} & \textbf{Memory (GB)} \\ \midrule
KNN & 35.24 & 100  & 0.76 \\
IVF & 8.39 & 98.45 & 0.78 \\
ANNOY & 11.83 & 98.39 & 1.2 \\
\rowcolor{lightgray} HNSW & 1.80 & 98.11 & 0.84 \\ \bottomrule
\end{tabular}
\label{table:chapter4-ann-all}
\end{table}

\subsection{User Study}

\begin{figure}[t!]
    \centering
    \includegraphics[width=\linewidth]{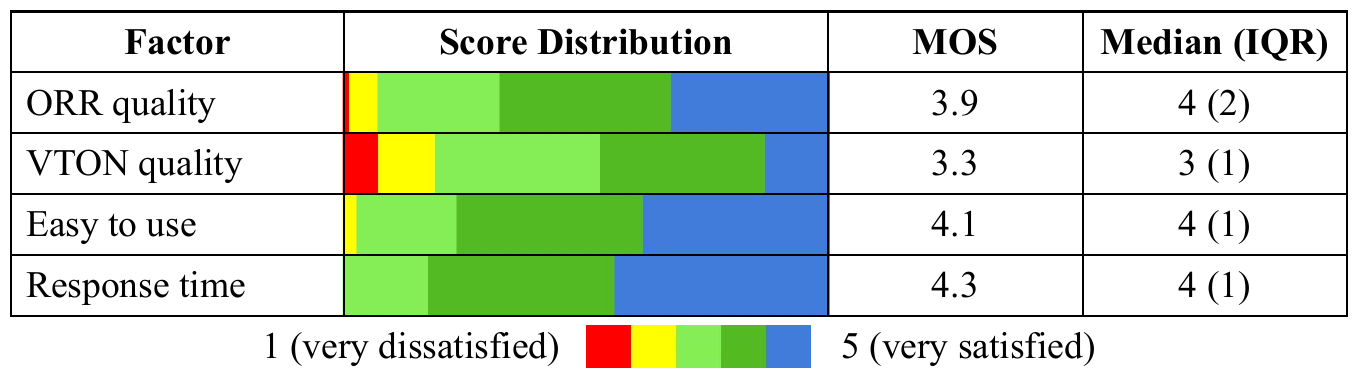}
    \caption{User study results measuring satisfaction of participants in our system 1: very dissatisfied, 5: very satisfied). MOS stands for Mean Opinion Score and IQR indicates Interquartile Range.}
    \label{fig:user-overall}
    \vspace{-3mm}
\end{figure}

\begin{figure}[t!]
\centering
    \includegraphics[width=\linewidth]{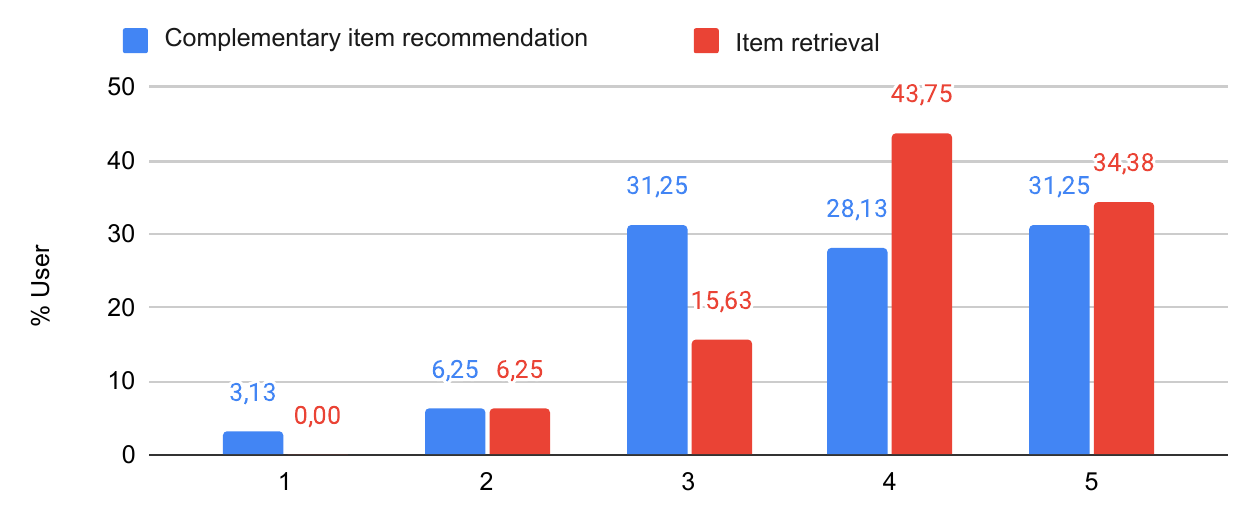}
    \caption{Rating scores on ORR quality (1: very dissatisfied, 5: very satisfied).}
    \label{fig:rec-result}
    \vspace{-5mm}
\end{figure}

We deployed our system for online users to try out and provide feedback. Our survey primarily aimed to measure users' satisfaction in four key factors: ORR quality, VTON quality, ease of use, and response time. \autoref{fig:user-overall} shows the average rating scores of each factor on a scale of 1 to 5 (1: very dissatisfied, 5: very satisfied) from the user study evaluation, which indicates that those users had a positive overall experience with our system.

Response times received the highest rating among the four factors, followed by ease of use.  Almost all users expressed their satisfaction with our friendly interface and processing speed. The seamless and fast visual search ensures that desired items are promptly displayed without delays. VTON factor also received positive feedback, highlighting the superb VTON quality, as it effectively assisted users in visualizing the garments in action, providing them with a valuable tool for making fashion choices.

As for the ORR results, the rating scores in \autoref{fig:rec-result} indicated users' satisfaction with them, highlighting the positive impact of item retrieval and item recommendation on enhancing the overall user experience. This feature helped users receive recommendations that align with their unique tastes and preferences, making their shopping journey more enjoyable and efficient.

Moreover, the user study revealed that 84,4\% of users found our system highly useful and effectively improved their online shopping experience. Furthermore, 65,6\% of the users expressed their willingness to recommend our system to their friends and family, emphasizing the positive reception and satisfaction among users.

\section{Conclusion}

In this paper, we develop a novel KiseKloset system supporting outfit retrieval, recommendation, and try-on. We explore ways to retrieve intra-category similar items and text feedback-guided items by utilizing the CLIP model. We also propose a novel transformer-based network to tackle inter-category complementary item recommendation issues. For practical usage, the used VTON is a parser-free lightweight network to boost processing speed while maintaining high-quality output and computational efficiency. Through intensive experiments, we prove the applicability and efficiency of our system in enhancing users' online shopping experience.

Our system focuses exclusively on upper-body garments. We intend to explore outfits belonging to other categories, such as shorts and pants (lower-body) or full-body dresses.

\section*{CRediT Authorship Contribution Statement}

\begin{itemize}
    \item \textbf{Thanh-Tung Phan-Nguyen}: Methodology, Conceptualization, Writing – original draft.
    \item \textbf{Khoi-Nguyen Nguyen-Ngoc}: Methodology, Visualization, Writing – original draft.
    \item \textbf{Tam V. Nguyen}: Validation, Writing – Review \& Editing.
    \item \textbf{Minh-Triet Tran}: Conceptualization, Writing – Review \& Editing.
    \item \textbf{Trung-Nghia Le}: Supervision, Conceptualization, Writing – Review \& Editing, Project administration.
\end{itemize}


\printcredits
\bibliographystyle{cas-model2-names}
\bibliography{sample-base}

@misc{Fashion2327-Statista2023,
    author       = {Statista},
	title        = {Fashion e-commerce market value worldwide from 2023 to 2030},
	note         = {Accessed: 2026-02-21},
	howpublished = {\url{https://www.statista.com/statistics/1298198/market-value-fashion-ecommerce-global/}},
}

@article{Vaswani-NeurIPS2017-Attention,
	title        = {Attention is all you need},
	author       = {Vaswani, Ashish and Shazeer, Noam and Parmar, Niki and Uszkoreit, Jakob and Jones, Llion and Gomez, Aidan N and Kaiser, {\L}ukasz and Polosukhin, Illia},
	year         = 2017,
	journal      = {NeurIPS},
	@pages        = {6000–6010}
}

@article{Devlin-ArXiv2018-BERT,
	title        = {BERT: Pre-training of Deep Bidirectional Transformers for Language Understanding},
	author       = {Devlin, Jacob and Chang, Ming-Wei and Lee, Kenton and Toutanova, Kristina},
	year         = 2018,
	journal      = {arXiv preprint arXiv:1810.04805},
}

@article{Radford-OpenAIblog2019-Language,
	title        = {Language models are unsupervised multitask learners},
	author       = {Radford, Alec and Wu, Jeffrey and Child, Rewon and Luan, David and Amodei, Dario and Sutskever, Ilya and others},
	year         = 2019,
	journal      = {OpenAI blog},
	volume       = 1,
	number       = 8,
	@pages        = 9,
}

@inproceedings{Baldrati-CVPR2022-Effective,
	title        = {Effective Conditioned and Composed Image Retrieval Combining CLIP-Based Features},
	author       = {Baldrati, Alberto and Bertini, Marco and Uricchio, Tiberio and Del Bimbo, Alberto},
	year         = 2022,
	booktitle    = {CVPR},
	@pages        = {21466--21474},
}

@inproceedings{Baldrati-CVPR2022-Conditioned,
	title        = {Conditioned and Composed Image Retrieval Combining and Partially Fine-Tuning CLIP-Based Features},
	author       = {Baldrati, Alberto and Bertini, Marco and Uricchio, Tiberio and Del Bimbo, Alberto},
	year         = 2022,
	booktitle    = {CVPR},
	@pages        = {4959--4968},
}

@inproceedings{Mariya-ECCV18-Learning,
	title        = {Learning Type-Aware Embeddings for Fashion Compatibility},
	author       = {Mariya I. Vasileva and Bryan A. Plummer and Krishna Dusad and Shreya Rajpal and Ranjitha Kumar and David Forsyth},
	year         = 2018,
	booktitle    = {ECCV},
	@pages        = {405--421},
}

@inproceedings{Lin-CVPR2020-Fashion,
	title        = {Fashion Outfit Complementary Item Retrieval},
	author       = {Yen-Liang Lin and Son Tran and Larry Davis},
	year         = 2020,
	booktitle    = {CVPR},
	@pages        = {3311--3319},
}

@inproceedings{Nguyen-ISMAR2023-DMVTON,
  title={DM-VTON: Distilled Mobile Real-time Virtual Try-On},
  author={Nguyen-Ngoc, Khoi-Nguyen and Phan-Nguyen, Thanh-Tung and Le, Khanh-Duy and Nguyen, Tam V and Tran, Minh-Triet and Le, Trung-Nghia},
  booktitle={ISMAR-Adjunct},
  @pages={695--700},
  year={2023},
}

@inproceedings{tangseng2017recommending,
  title={Recommending outfits from personal closet},
  author={Tangseng, Pongsate and Yamaguchi, Kota and Okatani, Takayuki},
  booktitle={ICCV Workshops},
  @pages={2275--2279},
  year={2017}
}

@inproceedings{jang2024lost,
  title={Lost Your Style? Navigating with Semantic-Level Approach for Text-to-Outfit Retrieval},
  author={Jang, Junkyu and Hwang, Eugene and Park, Sung-Hyuk},
  booktitle={WACV},
  @pages={8066--8075},
  year={2024}
}

@article{wu2022design,
  title={Design and implementation of virtual fitting system based on gesture recognition and clothing transfer algorithm},
  author={Wu, Ying and Liu, Hongbing and Lu, Pengzhen and Zhang, Lihua and Yuan, Fangjian},
  journal={Scientific Reports},
  volume={12},
  number={1},
  @pages={18356},
  year={2022},
  @@publisher={Nature Publishing Group UK London}
}

@inproceedings{chen2019pog,
  title={POG: personalized outfit generation for fashion recommendation at Alibaba iFashion},
  author={Chen, Wen and Huang, Pipei and Xu, Jiaming and Guo, Xin and Guo, Cheng and Sun, Fei and Li, Chao and Pfadler, Andreas and Zhao, Huan and Zhao, Binqiang},
  booktitle={SIGKDD},
  @pages={2662--2670},
  year={2019}
}

@inproceedings{liu2012hi,
  title={Hi, Magic Closet, Tell Me What to Wear!},
  author={Liu, Si and Tam V. Nguyen and Feng, Jiashi and Meng Wang and Yan, Shuicheng},
  booktitle={ACM Multimedia},
  @pages={619--628},
  year={2012}
}

@incollection{zhou2012image,
  title={Image-based clothes animation for virtual fitting},
  author={Zhou, Zhenglong and Shu, Bo and Zhuo, Shaojie and Deng, Xiaoming and Tan, Ping and Lin, Stephen},
  booktitle={SIGGRAPH Asia},
  @pages={1--4},
  year={2012}
}

@inproceedings{Sarkar-CVPRW2022-OutfitTransformer,
	title        = {OutfitTransformer: Outfit Representations for Fashion Recommendation},
	author       = {Sarkar, Rohan and Bodla, Navaneeth and Vasileva, Mariya and Lin, Yen-Liang and Beniwal, Anurag and Lu, Alan and Medioni, Gerard},
	year         = 2022,
	booktitle    = {CVPR Workshops},
	@pages        = {2262--2266},
}

@inproceedings{Han-ECCV2022-FashionViL,
	title        = {FashionViL: Fashion-Focused Vision-and-Language Representation Learning},
	author       = {Han, Xiao and Yu, Licheng and Zhu, Xiatian and Zhang, Li and Song, Yi-Zhe and Xiang, Tao},
	year         = 2022,
	booktitle    = {ECCV},
	@pages        = {634–651},
}

@article{Jegou-TPAMI2010-Product,
	title        = {Product quantization for nearest neighbor search},
	author       = {Jegou, Herve and Douze, Matthijs and Schmid, Cordelia},
	year         = 2010,
	journal      = {IEEE T-PAMI},
	volume       = 33,
	number       = 1,
	@pages        = {117--128},
}

@inproceedings{Sivic-ICCV2003-Video,
	title        = {Video Google: A text retrieval approach to object matching in videos},
	author       = {Sivic and Zisserman},
	year         = 2003,
	booktitle    = {ICCV},
	@pages        = {1470--1477},
}

@article{Malkov-TPAMI2018-Efficient,
	title        = {Efficient and robust approximate nearest neighbor search using hierarchical navigable small world graphs},
	author       = {Malkov, Yu A and Yashunin, Dmitry A},
	year         = 2018,
	journal      = {IEEE T-PAMI},
	volume       = 42,
	number       = 4,
	@pages        = {824--836},
}

@inproceedings{Liu-CVPR2016-DeepFashion,
	title        = {DeepFashion: Powering Robust Clothes Recognition and Retrieval with Rich Annotations},
	author       = {Liu, Ziwei and Luo, Ping and Qiu, Shi and Wang, Xiaogang and Tang, Xiaoou},
	year         = 2016,
	booktitle    = {CVPR},
	@pages        = {1096--1104},
}

@article{Lai-Arxiv2019-Contrastive,
	title        = {Contrastive Predictive Coding Based Feature for Automatic Speaker Verification},
	author       = {Lai, Cheng-I},
	year         = 2019,
	journal      = {arXiv preprint arXiv:1904.01575
        
        
        
        
        
        
        
        
        
        
        
        
        
        
        
        },
}

@inproceedings{He-CVPR2022-Style,
	title        = {Style-based global appearance flow for virtual try-on},
	author       = {He, Sen and Song, Yi-Zhe and Xiang, Tao},
	year         = 2022,
	booktitle    = {CVPR},
	@pages        = {3470--3479},
}

@inproceedings{Sandler-CVPR2018-Mobilenetv2,
	title        = {Mobilenetv2: Inverted residuals and linear bottlenecks},
	author       = {Sandler, Mark and Howard, Andrew and Zhu, Menglong and Zhmoginov, Andrey and Chen, Liang-Chieh},
	year         = 2018,
	booktitle    = {CVPR},
	@pages        = {4510--4520},
}

@inproceedings{Wu-CVPR2021-FashionIQ,
	title        = {Fashion IQ: A new dataset towards retrieving images by natural language feedback},
	author       = {Wu, Hui and Gao, Yupeng and Guo, Xiaoxiao and Al-Halah, Ziad and Rennie, Steven and Grauman, Kristen and Feris, Rogerio},
	year         = 2021,
	booktitle    = {CVPR},
	@pages        = {11307--11317},
}

@article{johnson-bigdata2019-faiss,
	title        = {Billion-scale similarity search with {GPUs}},
	author       = {Johnson, Jeff and Douze, Matthijs and J{\'e}gou, Herv{\'e}},
	year         = 2019,
	journal      = {IEEE Transactions on Big Data},
	volume       = 7,
	number       = 3,
	@pages        = {535--547},
}

@misc{web-recommendation-motivation,
    author       = {Visenze},
	title        = {Ace Your Product Recommendations to Grow Revenue},
	note         = {Accessed: 2023-07-19},
	howpublished = {\url{https://www.visenze.com/blog/2023/07/19/ace-your-product-recommendations-to-grow-revenue}},
}

\end{document}